\documentstyle[amscd,amssymb,verbatim,12pt]{amsart}
\def\dspace{\baselineskip=0.3 in}
\begin{document}
\dspace
\title[$f(R)$-COSMOLGY FROM EARLY ...  ]{$f(R)$-COSMOLGY FROM EARLY
  UNIVERSE UPTO FUTURE COSMIC COLLAPSE AND ITS AVOIDANCE}

\author{\bf S.K.Srivastava}
{ }
\maketitle
\centerline{ Department of Mathematics,}
 \centerline{ North Eastern Hill University,}
 \centerline{  Shillong-793022, India}
\centerline{ srivastava@@.nehu.ac.in; sushilsrivastava52@@gmail.com}

\vspace{1cm}

\centerline{\bf Abstract}
In the present model, a unified picture of cosmology from early inflation to late acceleration is obtained from $ f(R)-$ gravity with non-linear terms $ R^2 $ and $ R^5 $ of scalar curvature $R$. It is discussed that elementary particles and radiation are produced during early inflation. The emitted radiation thermalizes the universe to a very high temperature $ \sim 10^{18}{\rm GeV}$.  The exit of the universe from inflationary phase is followed by deceleration due to radiation-dominance heralding the standard cosmology with background radiation having the initial temperature $ \sim 10^{18}{\rm GeV}$. It is found that dark matter is induced by curvature and baryonic matter is produced during inflation. Radiation-dominated phase is followed by deceleration due to matter-dominance. The curvature-induced phantom dark energy dominates at the red-shift $ z = 0.303 $ causing the late acceleration. It is found that the universe will collapse in future. Further, it is investigated that back-reaction of quantum particles produced near the collapse time can avoid cosmic collapse and the universe will escape to revival of the state of early universe.


\vspace{1.6cm}

\centerline {\underline{\bf 1. Introduction}}
 \smallskip

Results of astrophysical observations, during last ten years, have conclusive
evidence for the late cosmic acceleration \cite{sp, ag}. It is caused by some exotic fluid
with negative pressure $p < - 1/3\rho_{\rm de}$, where
 $\rho_{\rm de} > 0$ is the density for the dark energy fluid. This condition violates the cosmic strong energy condition
(SEC) or weak energy condition (WEC). 

Many
field-theoretical and hydro-dynamical models of DE were proposed in the recent
past to explain the challenging observation of  late acceleration. A detailed
review of these models is available in
\cite{ejc}.In this race, curvature was also used as an important candidate and 
various models were proposed, where non-linear curvature terms were considered as gravitational alternatives of DE. These $f(R)-$ DE models are reviewed in
\cite{snj}. Recently, these models specially the model with non-linear terms $ R^n $ and $ R^{-m} $ \cite{sn03} received criticism in \cite{lds} on the ground
that these do not produce matter in
the late universe needed for formation of large scale structure in the
universe. In another review, Nojiri and Odintsov have discussed dark matter in
the late universe \cite{sn08} responding to this criticism \cite{lds}.

Here, we have a deviation in approach to obtain DE from curvature. In this approach, which
 was used earlier in refs.\cite{sks, sks06, sks07,sks08a, sks08b, sks08c}. It is contrary to the approach
 in refs.\cite{snj,lds, sn08}. Here, non-linear curvature terms are not assumed as DE like refs.\cite{snj,lds, sn08}. Rather, DE terms emerge from curvature spontaneously in this approach.  In
 refs.\cite{snj,lds, sn08}, gravitational equations are derived from the action having Einstein
-Hilbert term and non-linear curvature terms. Terms in gravitational equations, due to non
-linear curvature terms in the action, are recognized as DE terms. In the present paper as
 well as refs.\cite{sks, sks06, sks07,sks08a, sks08b, sks08c}, trace of $f(R)-$gravitational
 equations are obtained yielding an equation for scalar curvature $R$. In the homogeneous
 space-time, the equation for $R$ reduces to the second-order equation for the scale factor.
 First integral of this differential equation yields the Friedmann equation giving dynamics
 of the universe. Here, it is found that DE terms emerge in the Friedmann equation so
 derived.   
Thus, in the present model, DE terms are induced by linear as well as non-linear terms of
 scalar curvature, whereas only non-linear terms of curvature contribute to the dark energy in \cite{snj,lds, sn08}. 

Although the present model is based on $ f(R)-$ gravity, it is different from models in
 \cite{snj,lds, sn08} in two ways. The main difference is the difference in approach
 mentioned above. The other difference is the advantage of getting matter term from the gravitational sector. Moreover, there are many other interesting
 features of the present model. In the present set-up, non-linear terms of curvature are taken as $ R^2 $ and $ R^{(2 + r)} $ along with Einstein-Hilbert term. This paper is an extension of the work \cite{sks08b}. Here many other aspects of cosmology such as creation of SM particles in the early universe, collapse of the universe in future and possible avoidance of future cosmic collapse using quantum gravity are addressed to. In what follows, it is found that quintessence dark energy and dark radiation are induced by curvature in the early universe as well as dark matter and phantom dark energy are induced in the late universe if $r = 3$. Here, the topology of the cosmological model is given by homogeneous and flat FRW universe.
 
 The present model yields an interesting cosmological picture from the early universe to the future universe, which is given as follows. Here, investigations begin from the Planck scale, which is the fundamental scale. 
 It is found that the early universe inflates for a short
 period, driven by curvature-induced quintessence scalar behaving as an inflaton. During this period, elementary particles are produced and lot of energy is released
. The emitted energy thermalizes the universe rapidly upto the temperature $ \sim 10^{18}{\rm GeV}$. As a consequence, produced particles due to the decay of curvature-induced quintessence attain the thermal equilibrium with radiation. There are two sources of radiation (i) emitted radiation during inflation and (ii) curvature-induced radiation  recognized as dark radiation. The density of the emitted radiation during inflation dominates, so the universe is driven by emitted radiation and  decelerates after exit from inflation. Thus, in the present scenario, the standard cosmology is recovered at this epoch . The emitted radiation, during inflation, is identified with the cosmic  background radiation with very high initial temperature $ \sim 10^{18}{\rm GeV}$.

 The produced elementary particles, during inflation, undergo various processes of the standard cosmology such as nucleosynthesis,baryosynthesis and hydrogen-recombination, which are not discussed here. Thus, like radiation, we have two types of pressureless matter (i) 
baryonic matter, which is formed due to nucleosynthesis and baryosynthesis of elementary particles produced due to the decay of curvature inspired inflaton and (ii) curvature-induced non-baryonic matter identified as dark matter. After sufficient expansion of the universe, the matter dominates over radiation causing decelerated expansion as $ \sim t^{2/3} $ in the late universe. 

 It is interesting to note that, in the late universe, curvature-induced phantom terms in the Friedmann equation are $ \rho_{\rm ph} [1 - \rho_{\rm ph}/2\lambda] $ with $ \rho_{\rm ph} $ being the phantom energy density. Moreover, curvature causes another constant $ \lambda $, which is analogous to negative brane-tension in Randall-Sundrum II theory of brane-gravity \cite{rs1, rs2, rm}. As a remark, it is nice to mention that these type of terms also appear in Freidmann equation based on the loop quantum gravity \cite{ms}. As brane-theory prescriptions are not used here, the curvature-induced $ \lambda $ appearing in this model is identified as {\em cosmic tension} like the Refs. \cite{sks08b, sks08c}.

Further, it is found that due to continuous increase in phantom energy density in the expanding universe, dominance of curvature-induced phantom begins at the red-shift $ z = 0.303 $ causing a cosmic jerk. As a consequence, a transition from deceleration to acceleration takes place in the very late universe. It is found that the phenomenon of cosmic acceleration will continue in future too.

In the dynamics of the future universe, the cosmic tension plays a very important role as it opposes expansion of the universe. This effect is significant on sufficient increase in phantom energy density. Due to growth 
in phantom energy density in the expanding universe, it is found that matter will re-dominate cosmic dynamics, when phantom energy density will grow to twice of the brane-tension. As a consequence, universe will decelerate. 

Increase in $ \rho_{\rm ph} $ will still continue with the increasing scale factor $ a(t) $. So, phantom terms  in the Friedmann equation will be negative when $ \rho > 2\lambda $. It is found that expansion will stop on the growth of the scale factor upto a maximum value $ a_m = 1.88\times 10^{10} $ at time $ t_m \simeq 3.45\times 10^{15} t_0 $ ($ t_0 $ being the present age of the universe) and the universe will bounce causing contraction in the universe. As a result, matter energy density will increase rapidly such that energy density and pressure density diverge as well as $ a = 0 $ at $ t_{\rm col} \simeq 3.62\times 10^{15} t_0 $. It means that big-crunch singularity will be caused and the universe will collapse. This result is based on the classical mechanics. 

It is imperative to note that energy density ,pressure and curvature will be extremely high near the time of the cosmic collapse $ t_{\rm col} $. So, large structures of the universe will get smashed to elementary particles. This state is analogous to the early universe. So, quantum gravity effects can not be ignored near $ t_{\rm col} $ as quantum effects are important in the early universe. As a result, quantum particles will get produced due to rapid change in topology of the space-time. It is found that cosmic collapse can be avoided due to the back-reaction of these particles and the universe will expand exponentially after $ t > t_{\rm col} $. Thus, this model predicts revival of the state of early universe in future when $ t > t_{\rm col} $.

The paper is organized as follows. In section 2,$f(R)$-gravitational equations are derived and trace of these equations are obtained. Curvature-inspired Friedmann equation is obtained in the early and late universe are obtained from the trace of gravitational equations.  The early universe is discussed in section 3. In this section, it is found that curvature-induced dark energy mimics quintessence and the early universe undergoes the power-law inflation. During curvature-induced quintessence scalar behaves as an inflaton. Thus, the source of inflaton is known, which is an important consequence of this model. Further, it is discussed that the curvature-inspired inflaton decays during this period. This decay causes production of elementary particles and radiation, which thermalizes the early universe upto a very high temperature $ \sim 10^{18}{\rm GeV}$. In section 4, it is discussed that the standard cosmology is recovered after exit from inflation and decelerates being driven by radiation and subsequently by matter. This section also contains transition from deceleration to acceleration driven by curvature-induced phantom. Re-dominance of matter in future, cosmic collapse in future and its possible avoidance using the quantum  gravity are addressed in section 5. In the last section, salient features of the present model and concluding remarks are given.

Natural units$(k_B = {\hbar} = c = 1)$ (where $k_B, {\hbar}, c$ have their
usual meaning) are used here. GeV is used as a fundamental unit and we have $1 {\rm GeV}^{-1}
= 6.58 \times 10^{-25} sec$ and $1 {\rm GeV} = 1.16 \times {10^{13}}^0 K.$

\bigskip

\centerline {\underline{\bf 2.  $f(R)$- gravity and Friedmann equations }}

 \smallskip
 The action is taken as
$$ S = \int {d^4x} \sqrt{-g} \Big[\frac{R}{16 \pi G} + \alpha R^2 + \beta
R^{(2 + r)}  \Big],  \eqno(2.1)$$ 
where $G = M_P^{-2} (M_P = 10^{19} {\rm GeV}$ is the Planck mass), $\alpha$ is
a dimensionless coupling constant, $\beta$ is a constant having dimension
(mass)$^{(-2 r)}$ (as $R$ has mass dimension 2) with $r$ being a positive real
number.

Using the condition $\delta S/\delta g^{\mu\nu} = 0$, the action (2.1) yields gravitational field equations
$$ \frac{1}{16 \pi G} (R_{\mu\nu} - \frac{1}{2} g_{\mu\nu} R) + \alpha (2
\triangledown_{\mu} \triangledown_{\nu} R - 2 g_{\mu\nu} {\Box} R -
\frac{1}{2} g_{\mu\nu} R^2 + 2 R R_{\mu\nu} ) $$
$$ + \beta (2 + r) ( \triangledown_{\mu} \triangledown_{\nu} R^{(1 + r)} -
g_{\mu\nu} {\Box} 
R^{(1 + r)}) + \frac{1}{2}\beta g_{\mu\nu} R^{(2 + r)} $$
$$- \beta (2 + r) R^{(1 +   r)} R_{\mu\nu}   = 0, \eqno(2.2)$$  
where $\triangledown_{\mu}$ stands for the covariant derivative.

Taking trace of (2.2), it is obtained that
$$ - \frac{R}{16 \pi G} - 6 \alpha{\Box} R - 3 \beta (2 + r) {\Box} R^{(1 +
 r)}  + \beta r R^{(2 + r)}  = 0 \eqno(2.3)$$
with
$$ {\Box} = \frac{1}{\sqrt{-g}} \frac{\partial}{\partial x^{\mu}}
\Big(\sqrt{-g} g^{\mu\nu} \frac{\partial}{\partial x^{\nu}} \Big). \eqno(2.4)$$

In (2.3)
$${\Box} R^{(1 + r)} = (1 + r) [R^{r}{\Box} R + r
R^{(r - 1)} \triangledown^{\mu}R \triangledown_{\mu}R ]. \eqno(2.5)$$

From (2.3) and (2.5)
$$ - \frac{R}{16 \pi G} - [6 \alpha + 3 \beta (1 + r)(2 + r)  R^r ] {\Box} R -
3 \beta r (1 + r)(2 + r)R^{(r - 1)} \triangledown^{\mu}R \triangledown_{\mu}R$$
$$+ \beta r R^{(2 + r)}  = 0 \eqno(2.6)$$ 

In (2.6), $[6 \alpha + 3 \beta (1 + r)(2 + r)  R^r ]$
emerges as a coefficient of ${\Box} R$ due to 
presence of terms $\alpha R^2$ and $\beta R^{(2 + r)}$ in the action
(2.1). If $\alpha = 
0$, effect of $R^2$ vanishes and  effect of $R^{(2 + r)}$ is switched
off for $\beta = 0$. So, an {\em effective} scalar curvature
${\tilde R}$ is defined as \cite{sks07} 
$$ \gamma {\tilde R}^r = [6 \alpha + 3 \beta (1 + r)(2 + r)  R^r ] , \eqno(2.7)$$
where  $\gamma$ is a constant having dimension (mass)$^{-2r}$ being
used for dimensional correction.

Using (2.7) in (2.6), we have
$$\frac{1}{16 \pi G} Y^{1/r} - (\gamma/r) {\tilde R}^r Y^{(1/r - 2)} [Y {\Box}
Y + (1/r - 1) 
\triangledown^{\mu} Y \triangledown_{\mu} Y ]$$ 
$$ - 3 (\beta/r) (1 + r)(2 + r) Y^{(1/r - 1)} \triangledown^{\mu} Y \triangledown_{\mu} Y  + \beta r Y^{(2 + r)/r} = 0, \eqno(2.8)$$
where
$$ Y = R^{r} = \frac{ \gamma {\tilde R}^{r} - 6 \alpha}{3 \beta (1 + r)(2 + r)}. \eqno(2.9a)$$
(2.8) is simplified as 
 $$\frac{1}{16 \pi G} Y - (\gamma/r) {\tilde R}^{r}[{\Box} Y + (1/r - 1)Y^{-1} \triangledown^{\mu} Y \triangledown_{\mu} Y ]$$
$$ - 3 (\beta/r) (1 + r)(2 + r)  \triangledown^{\mu} Y \triangledown_{\mu} Y + \beta r Y^{(1/r + 2)} = 0. \eqno(2.9b) $$

Using (2.9a)in (2.9b), it is obtained that
$$-\frac{r}{16 \pi G \gamma} \Big[\frac{6 \alpha}{\gamma {\tilde R}^r} - 1 \Big] + {\Box}{\tilde R}^r - (1/r - 1) \frac{\gamma}{[6 \alpha - \gamma {\tilde R}^{r}]} \triangledown^{\mu}{\tilde R}^{r} \triangledown_{\mu}{\tilde R}^{r} + {\tilde R}^{- r} \triangledown^{\mu}{\tilde R}^{r}\triangledown_{\mu}{\tilde R}^{r}$$
$$ + [3 \beta^2 r (1 + r)(2 + r)/ \gamma^2] {\tilde R}^{r} \Big[\frac{\gamma {\tilde R}^{r}- 6 \alpha}{ 3 \beta (1 + r)(2 + r)} \Big]^{(1/r + 2)} = 0 . \eqno(2.10)$$

(2.10) is re-written as
$$-\frac{1}{16 \pi G} \frac{1}{\gamma {\tilde R}^{r - 1}}\Big[\frac{6
  \alpha}{\gamma {\tilde R}^{r}} - 1 \Big] + {\Box}{\tilde R} +
  (r - 1) {\tilde R}^{-1} \triangledown^{\mu}{\tilde R}
  \triangledown_{\mu}{\tilde R}$$
$$ - (1 - r) \frac{\gamma {\tilde R}^{r - 1}}{6 \alpha - \gamma {\tilde R}^{r}}\triangledown^{\mu}{\tilde R}\triangledown_{\mu}{\tilde R} +
  r {\tilde R}^{-1} \triangledown^{\mu}{\tilde R} 
  \triangledown_{\mu}{\tilde R}$$
$$+ r)(2 + r)/ \gamma^2]
  {\tilde R}^{2 r - 1} \Big[\frac{\gamma {\tilde R}^{r} - 6\alpha}{3 \beta (1 + r)(2 + r)} \Big]^{(1/r + 2)} = 0 . \eqno(2.11)$$ 

Experimental evidences \cite{ad} support spatially homogeneous and
flat model of the universe 
$$dS^2 = dt^2 - a^2(t) [dx^2 + dy^2 + dz^2] \eqno(2.12)$$
with $a(t)$ being the scale factor.

For $a(t)$, being the power-law function of cosmic time, ${\tilde R} \sim a^{-n}$. For
example,  ${\tilde R} \sim a^{-3}$ for
matter-dominated model. So, there is no harm in taking
$$  {\tilde R} = \frac{A}{ a^n} , \eqno(2.13)$$
where $n > 0$ is a real number and $A$ is a constant with mass dimension 2. 

Connecting (2.11) and (2.13), it is obtained that
$$ \frac{\ddot a}{a} + \Big[2 - n - n (r - 1) + \frac{n (1 - r) \gamma A^r
  a^{- nr}}{6 \alpha - \gamma A^r a^{- nr}} - n r \Big] \Big(\frac{\dot a}{a} \Big)^2 = \frac{a^{n r}}{16 \pi G \gamma A^r} \Big[\frac{6 \alpha a^{nr}}{\gamma A^r} - 1 \Big]$$
$$ - \frac{ \beta^{-1/3} }{n (\gamma A^r a^{-nr})^2[3r(1 + r)(2 + r)]^{1 + 1/r} }[6 \alpha - \gamma A^r a^{-nr}]^{2 + 1/r}  \eqno(2.14)$$
taking $(- \beta)^{-1/3} = - \beta^{-1/3}$ and ignoring complex roots as these roots lead  to unphysical situations. Now, we have following two cases.

\smallskip

\noindent {\bf Case 1 : The Early Universe}

In this case, $a(t)$ is very small, so (2.14) is approximated as
$$ \frac{\ddot a}{a} + \Big[2 - n - n r \Big] \Big(\frac{\dot a}{a} \Big)^2 \simeq - \frac{\beta^{-1/3} }{n (\gamma A^r a^{-nr})^2[3r(1 + r)(2 + r)]^{1 + 1/r} }$$
$$\times[6 \alpha - \gamma A^r a^{-nr}]^{2 + 1/r}  \eqno(2.15)$$
as
$$\frac{ \gamma A^r
  a^{- nr}}{6 \alpha - \gamma A^r a^{- nr}} \approx - 1.\eqno(2.16)$$

Integration of (2.16) leads to
$$ \Big(\frac{\dot a}{a} \Big)^2 = \frac{B}{a^{(2 + 2M)}} - \frac{2 \beta^{-1/r}}{n (\gamma A^r)^2 [3r(1 + r)(2 + r)]^{1 + 1/r}a^{(2 + 2M)}}$$
$$\times \int a^{(1 + 2M + 2nr)}[6 \alpha - \gamma A^r a^{-nr}]^{2 + 1/r}  \eqno(2.17)$$
with
$$ M = 2 - n - n r. \eqno(2.18)$$

\smallskip
\noindent {\bf Case 2 : The Late Universe}

In this case, $a(t)$ is large, so (2.14) is approximated as
$$\frac{\ddot a}{a} + \Big[2 - n - n (r - 1)- nr \Big] \Big(\frac{\dot a}{a} \Big)^2 \simeq  \frac{a^{n r}}{16 \pi G \gamma A^r} \Big[\frac{6 \alpha a^{nr}}{\gamma A^r} - 1 \Big] $$
$$ - \frac{ \beta^{-1/3} }{n (\gamma A^r a^{-nr})^2[3r(1 + r)(2 + r)]^{1 + 1/r} }(6\alpha)^{2 + 1/r} [a^{2 nr} - (2 + 1/r)\gamma A^r a^{nr}] \eqno(2.19a)$$
as 
$$\frac{ \gamma A^r
  a^{- nr}}{6 \alpha - \gamma A^r a^{- nr}} \approx 0$$
for large scale factor $a$. So, (2.19a) is re-written as
$$\frac{\ddot a}{a} + \Big[2 - 2nr \Big] \Big(\frac{\dot a}{a} \Big)^2 = D a^{nr} - E a^{2nr} , \eqno(2.19b)$$
where
$$D = \Big(\frac{6\alpha}{\gamma A^r}\Big) \Big[ \frac{1}{16 \pi G n} - (2 +
1/r)\frac{[3 r (1 + r)(2 + r)]^{-1-1/r}}{n}\Big(\frac{6\alpha}{\gamma
  A^r}\Big) \Big] \eqno(2.20a)$$  
and
$$E = \Big(\frac{6\alpha}{\gamma A^r}\Big)^2 \Big[ \frac{1}{16 \pi G n} -
\frac{[3 r (1 + r)(2 + r)]^{-1-1/r}}{n}\Big(\frac{6\alpha}{\gamma A^r}\Big)
\Big]. \eqno(2.20b)$$ 

(2.19b) is integrated to
$$ \Big(\frac{\dot a}{a} \Big)^2 = \frac{B}{a^{(2 + 2N)}} + \frac{2D}{(2 + 2N +
    nr)} a^{nr} \Big[1 - \frac{E(2 + 2N + nr)}{D(2 + 2N + 2nr)}a^{nr} \Big]
  \eqno(2.21)$$  
with
$$ N = 2 - 2nr .\eqno(2.22)$$ 

Further, it is found that if $M = 1$, the first term on r.h.s.(right hand
side) of (2.17)gives radiation. Moreover, if $N = 1/2$  the first term of r.h.s. of
(2.21) gives matter. So, setting $M = 1$ in (2.18) and $N = 1/2$ in (2.22) to get aviable model, it is
obtained that 
$$ nr = \frac{3}{4}, \eqno(2.23)$$
$$ n = \frac{1}{4}\eqno(2.24)$$
and
$$ r = 3. \eqno(2.25)$$

\bigskip

\centerline {\underline{\bf 3. Power-law inflation, particle creation and thermalization  }} 

\centerline {\underline{\bf in the early universe }}

 \smallskip

\noindent {\underline{\bf 3(a).Power-law inflation}} 
\smallskip

 The approximated Friedmann equation (2.17), in the case of the early universe, looks like
$$ \Big(\frac{\dot a}{a} \Big)^2 = \frac{B}{a^4} - \frac{8 \beta^{-1/3}}{ (\gamma A^3)^2 [180]^{4/3}a^4} \int a^{9/2}[6 \alpha - \gamma A^3 a^{- 3/4}]^{7/3} {da} \eqno(3.1)$$
using definitions of $M$ and $N$ as well as (2.24) and (2.25). In (3.1),
$$\int a^{9/2}[6 \alpha - \gamma A^3 a^{- 3/4}]^{7/3}{da} = \Big[\frac{2}{11} a^{11/2}\{6 \alpha - \gamma A^3 a^{- 3/4}\}^{7/3}\Big] - \frac{7}{22} \gamma A^3$$
$$\times \int a^{15/4}\{6 \alpha - \gamma A^3 a^{- 3/4}\}^{4/3} da .\eqno(3.2a)$$
It is noted that for
$$a < \Big(\gamma A^3/6 \alpha \Big)^{4/3} = a_c , \eqno(3.2b)$$
terms within bracket and the integral on the right hand side of (3.2a) are of
the order of $a^{15/4}$.

So,
$$\int a^{9/2}[6 \alpha - \gamma A^3 a^{- 3/4}]^{7/3} {da}\approx [\frac{2}{11}
a^{11/2}\{6 \alpha - \gamma A^3 a^{- 3/4}\}^{7/3} .\eqno(3.2c)$$ 
Thus, using (3.2a,b,c),(3.1) is approximated as
$$ \Big(\frac{\dot a}{a} \Big)^2 \approx \frac{B}{a^4} - \frac{16
  \beta^{-1/3}}{11 (\gamma A^3)^{-1/3} [180]^{4/3}}  a^{3/2}\Big[a_c^{-3/4} - a^{- 3/4}\Big]^{7/3}  \eqno(3.3)$$ 

It is interesting to see that a radiation density  term $B/a^4$ emerges
spontaneously. This type of term emerged first in brane-gravity inspired Friedmann
equation. So, analogous to brane-gravity, here also $B/a^4$ is identified as dark radiation. Other terms on r.h.s. of (3.3) are caused by linear as well as non-linear
terms of 
curvature in the action (2.1). These terms also constitute energy density term
$$ \rho^{\rm qu}_{\rm de} =  \frac{3}{8 \pi G} \Big[\frac{16
  \beta^{-1/3}}{11 (\gamma A^3)^{-1/3} [180]^{4/3}}  a^{3/2}\Big]\Big[ a^{- 3/4} -
  \frac{6 \alpha}{\gamma A^3}\Big]^{7/3}  \eqno(3.4)$$
(taking real root of $(-1)^{-1/3}$ as above) satisfying the conservation
  equation 
$$ {\dot \rho}_{\rm de} + 3 \frac{\dot a}{a} ( \rho_{\rm
  de} + p_{\rm de} ) = 0. \eqno(3.5)$$

Connecting (3.4) and (3.5),  equation of state (EOS)
is obtained as

$$p^{\rm qu}_{\rm de} =  - \frac{3}{2}\rho^{\rm qu}_{\rm de} + \frac{7}{12} f [a^{-3/4} -
a_c^{-3/4}]^{4/3}, \eqno(3.6a)$$
where
$$ f = \frac{3}{8\pi G}  \frac{16
  \beta^{-1/3}}{11 (\gamma A^3)^{-1/3} [180]^{4/3}}  \eqno(3.6b)$$
(3.6a) is the scale factor-dependent equation of state parameter,valid for $a_P
\le a(t) < a_c$. Such an equation of state parameter is obtained in
\cite{sks06} also. It yields
$$\rho^{\rm qu}_{\rm de} + p^{\rm qu}_{\rm de} > o \quad {\rm and}\quad
\rho^{\rm qu}_{\rm de} + 3 p^{\rm qu}_{\rm de} < o ,$$
for $a_P \le a(t) \le a_c$.
It shows that DE, having energy density (3.4) mimics {\em quintessence dark
  energy} \cite{sks, sks06,  sks07, sks08a, sks08b}.

Here investigations start at the Planck scale, where DE density is obtained
around $10^{75} {\rm GeV}^4$. So, (3.4) is re-written as
$$\rho^{\rm qu}_{\rm de} = F a^{3/2} [ a^{- 3/4} - a_c^{- 3/4}]^{7/3}
\eqno(3.7 a)$$
with
$$ F =  10^{75} {a_P^{-3/2}\Big[ a_P^{- 3/4} - a_c^{- 3/4} 
  \Big]^{-7/3}}.\eqno(3.7b)$$
Thus , (3.6b) and (3.7b) imply
$$ f = F .\eqno(3.7c)$$

(3.3) and (3.7a) imply
$$ \Big(\frac{\dot a}{a} \Big)^2 \simeq \frac{B}{a^4} + \frac{8 \pi \times
  10^{37}}{3} \Big(\frac{ a}{a_P}\Big)^{3/2} \Big[\frac{ a^{- 3/4} - a_c^{-
  3/4}  }{ a_P^{- 3/4} - a_c^{- 3/4}}
  \Big]^{7/3} \eqno(3.8)$$ 
using $ G = M_P^{-2} = 10^{-38} {\rm GeV}^{-2}.$

As $a_P$ is expected to be extremely small, so 
$\rho^{\rm qu}_{\rm de}$ dominates over the radiation term in (3.8).  Moreover,
(3.7a) shows that $\rho^{\rm qu}_{\rm de}$ vanishes at $a = a_c$. So, for $a_P <
a(t) < a_c,$ cosmic dynamics is given by
\begin{eqnarray*}
 \Big(\frac{\dot a}{a} \Big)^2 &\simeq & \frac{8 \pi \times
  10^{37}}{3} \Big(\frac{ a}{a_P}\Big)^{3/2} \Big[\frac{ a_c^{- 3/4} - a^{-
  3/4} }{ a_c^{- 3/4} - a_P^{- 3/4}}  \Big]^{7/3} \\ &\simeq & \frac{8 \pi \times
  10^{37}}{3} \Big(\frac{ a}{a_P}\Big)^{-1/4}. 
\end{eqnarray*}
\vspace{-1.2cm}
\begin{flushright}
(3.9) 
\end{flushright}
\smallskip

(3.9) integrates to
$$ a(t) = a_P \Big[1 + \frac{M_P}{8 \sqrt{3 \pi}}(t - t_P) \Big]^8
\eqno(3.10)$$ 
showing {\em acceleration} as ${\ddot a} > 0.$ 

If expansion (3.10) yields sufficient inflation in the early universe, 
$$ \frac{a_c}{a_P} = 10^{28}. \eqno(3.11)$$
The universe comes out of the inflationary phase at $t = t_c$ when $a(t)$
acquires the value $a_c$. So, from (3.10) and (3.11), it is obtained that
$$  t_c \simeq   7.77 \times 10^{4} t_P  \eqno(3.12)$$
using (3.11).

\smallskip

\noindent {\underline{\bf 3(b). Realization of curvature induced quintessence dark }}

\noindent {\underline{\bf energy through scalar field $\phi$ }} 
\smallskip

In 3(a), it is found that curvature induced quintessence dark energy causes
power-law inflation. Though, we have a gravitational origin of quintessence DE
here, it is natural to probe a scalar $\phi(t,{\bf x})$ giving
$\rho_{\rm de}$ obtained above using the scheme adapted in \cite{sks08a,jdb}. With $V(\phi)$ as potential and having minimal
coupling with gravity, in the homogeneous and flat model of the universe
(2.12), $\phi(t,{\bf x})$ obeys the
equation
$$ {\Box} \phi(t,{\bf x}) +  V^{\prime}(\phi) = 0 \eqno(3.13a)$$
and has mass dimension equal to 1. Here, $V^{\prime}(\phi) = d V/d\phi.$ 

In the homogeneous and flat model of the universe
(2.12), (3.13a) is obtained as
$${\ddot \phi_0} + 3 \frac{\dot a}{a}{\dot \phi_0} + V^{\prime}(\phi)_{\phi = \phi_0} = 0
\eqno(3.13b)$$
as $\phi(t,{\bf x}) = \phi_0(t)$ due to homogeneity. 
DE density and pressure for $\phi(t)$ are given as
$$ \rho_{\rm de} = \frac{1}{2} {\dot \phi_0}^2 + V(\phi_0) \eqno(3.14a)$$
and
$$ p_{\rm de} = \frac{1}{2} {\dot \phi_0}^2 - V(\phi_0) \eqno(3.14b)$$

It is interesting to see that  conservation equation (3.5) yields
(3.13b) for DE density and pressure of the $\phi_0(t)$- fluid 
given by (3.14a) and (3.14b), 

It is demonstrated above that quintessence DE, given by (3.7a) and (3.7b),
derives the power-law inflation with $a(t)$ given by (3.10). The Raychoudhuri
equation (which is obtained connecting Friedmann equation and conservation
equation) yields
$$ {\dot H} = - 4 \pi G ( \rho_{\rm de} + p_{\rm de}) = - 4 \pi G {\dot
  \phi_0}^2 \eqno(3.15)$$
using (3.14a) and (3.14b). $H$, in (3.15), is obtained from (3.10) as
$$ H = \frac{\dot a}{a} = \frac{M_P}{\sqrt{3 \pi}} \Big[1 + \frac{M_P}{8
  \sqrt{3 \pi}}(t - t_P) \Big]^{-1}.  \eqno(3.16)$$

(3.15) and (3.16) yield
$$ {\dot \phi_0} =  \frac{M^2_P}{4 \pi \sqrt{6}} \Big[1 + \frac{M_P}{8
  \sqrt{3 \pi}}(t - t_P) \Big]^{-1}.  \eqno(3.17)$$

(3.17) integrates to
\begin{eqnarray*}
 \phi_0 &=& M_P \sqrt{\frac{2}{\pi}} ln \Big[1 + \frac{M_P}{8
  \sqrt{3 \pi}}(t - t_P) \Big]\\ &=& M_P \sqrt{\frac{1}{32 \pi}} ln
  \Big(\frac{a}{a_P} \Big)
\end{eqnarray*}
\vspace{-1.8cm}
\begin{flushright}
(3.18)
\end{flushright}
using (3.10) and $\phi^P_0 = 0$ at Planck scale. Further, connecting (3.11) and (3.18), it is evaluated that
$$ \phi^c_0 = 6.43 M_P  \eqno(3.19)$$
being $ \phi_0$ at the end of inflation. (3.18) yields the relationship
$$ a(t) = a_P e^{[\phi_0 M_P^{-1} \sqrt{32 \pi} ]} .  \eqno(3.20)$$

As $\phi_0$ is the quintessence scalar, giving curvature induced quintessence
DE. So, using (3.6a,b) and (3.7a,b,c), it is obtained that
\begin{eqnarray*}
 V(\phi_0) &=& \frac{1}{2} (\rho_{\rm de} - p_{\rm de}) \\&=& \frac{5}{4} F
 [a^{-3/4} - a_c^{-3/4}]^{4/3} \Big[a^{3/2} ([a^{-3/4} - a_c^{-3/4}) -
 \frac{7}{30}\Big] 
\end{eqnarray*}
\vspace{-1.8cm}
\begin{flushright}
(3.21)
\end{flushright}
with $ a(t)$ given by (3.20). As $ V(\phi_0) = 0$ at $\phi_0 = \phi_0^c,$ it
 shows that quintessence scalar $\phi_0$ falls from the high hill of the
 potential to the ground state at the end of inflation. For $\phi_0 <
 \phi_0^c$, (3.21) is approximated as
$$  V(\phi_0) \simeq \frac{5}{4} F a_P^{-1/4}  e^{-[\phi_0 M_P^{-1} \sqrt{2 \pi}
]}   \eqno(3.22)$$
with $F$ given by (3.7b).
Thus, it is obtained that curvature induced quintessence DE driving power-law
inflation can be realized through quintessence scalar $\phi_0(t)  $ given by (3.20).

 In non-gravitational models of
DE, origin of quintessence is not known. This model has an advantage over
non-gravitational DE models due to gravitational origin of quintessence scalar.

\smallskip

\noindent {\underline{\bf 3(c). Particle creation during inflation}}

\smallskip

\noindent \underline{\bf Creation of spinless bosons}

$\phi_0(t)$ is the background field deriving inflation and $\phi(t,x)$ can be
realized as $ \phi(t,{\bf x}) = \phi_0(t) + \delta \phi(t,{\bf x})$ with $\delta
\phi(t,{\bf x})$ being the quantum fluctuation. Here, perturbations in the metric
components are ignored for simplicity. So, from (3.13a) and (3.13b), we
obtain
$$ {\Box}\delta \phi(t,{\bf x}) + + V^{\prime\prime}(\phi)_{\phi = \phi_0} \delta \phi(t,{\bf x}) = 0 \eqno(3.23a)$$

In the space-time (2.12), (3.23a) looks like
$$ {\ddot \delta\phi(t,{\bf x})} + 3 \frac{\dot a}{a}{\dot \delta\phi(t,{\bf x})} - a^{-2} \Big[
\frac{\partial^2}{\partial x^2} + \frac{\partial^2}{\partial y^2} +
\frac{\partial^2}{\partial z^2} \Big]\delta\phi (t, {\bf x})$$
$$ + V^{\prime\prime}(\phi)_{\phi = \phi_0} \delta \phi(t,{\bf x}) = 0 .\eqno(3.23b)$$

Using the decomposition
$$  \delta \phi(t, {\bf x}) = {\sum_{- \infty}^{\infty}} [\phi_k (t) a_k
e^{i {\vec k}.{\vec x}} + \phi^*_k (t) a_k^{\dag}
e^{-i {\vec k}.{\vec x}} ], \eqno(3.24)$$
$a(t)$ given by (3.10) and $V(\phi)$ given by (3.22), (3.23b) is obtained as
$$\frac{d^2 \phi_k}{d {\eta}^2} + \frac{24}{\eta} \frac{d \phi_k}{d {\eta}} +
\Big[\frac{192 \pi k^2}{M_P^2 \eta^{16}} + \frac{60}{\eta^2} \Big]\phi_k = 0 
\eqno(3.25a)$$ 
for $a < a_c$. Here
$$ \eta = \Big[1 + \frac{M_P}{8 \sqrt{3 \pi}}(t - t_P) \Big]. \eqno(3.25b)$$ 

 $\delta \phi(\eta, {\bf x})$ and $\Pi = {\partial \delta \phi(\eta, {\bf x})}{\partial {\eta}} $ in (3.24) satisfy quantum conditions
\begin{eqnarray*}
&&[ \delta \phi(\eta, {\bf x}) , \delta \phi(\eta, {\bf x}^{\prime})] = 0 \\&&[\Pi(\eta, {\bf x}) , \Pi(\eta, {\bf x}^{\prime})] = 0 \\&& [\delta \phi(\eta, {\bf x})  , \Pi(\eta, {\bf x}^{\prime}) ] =  i (-g)^{-1/2}
 \delta^3 ({\bf x} - {\bf x}^{\prime})
\end{eqnarray*}
\vspace{-2.5cm}
\begin{flushright}
(3.25c,d,e)
\end{flushright}
\vspace{0.3cm}
with $\delta^3 ({\bf x} - {\bf x}^{\prime})$ being the Dirac delta function.

(3.25a) integrates to
\begin{eqnarray*}
 \phi_k(t) &=& \phi_k(\eta) \\&=& \eta^{-23/2} \Big[C_1 J_{ir}(- k \sqrt{192
  \pi} \eta^{-7}/7 a_P M_P) \\&& + C_2 Y_{ir}(- k \sqrt{192 \pi} \eta^{-7}/7 a_P
  M_P) \Big] 
\end{eqnarray*}
\vspace{-1.8cm}
\begin{flushright}
(3.26a)
\end{flushright}
where $J_n(x) $ and  $Y_{n}(x) = (-1)^n J_n(x) $ are Bessel's functions  with
$$ n = \pm ir = \pm \frac{i}{7} \sqrt{\frac{60}{\sqrt{a_P}} - 11.5^2 } \eqno(3.26b)$$
with $i = \sqrt{-1}$. $C_1$ and $C_2$ are integration constants. Moreover, from (3.12) and (3.25b), it is obtained that $
1 \le \eta < 7.77 \times 10^4 t_P$. So, for $a_P \approx 10^{-47}$, which is
possible for $a_P$ being the scale factor at Planck scale,  $|k \sqrt{192 \pi}
\eta^{-7}/7 a_P M_P|$ is large.

Using identities
$$ J_n (x) \frac{d J_{-n}(x)}{dx} - \frac{d J_{n}(x)}{dx} J_{-n}(x) = -
\frac{2 sin (n \pi )}{\pi x} ,\eqno(3.26c)$$

$$ Y_n (x) \frac{d Y_{-n}(x)}{dx} - \frac{d Y_{n}(x)}{dx} Y_{-n}(x) = -
\frac{2 sin (n \pi )}{\pi x}, \eqno(3.26d)$$
and (A6),
$$  \phi^{(1)}_k(t) = C_1 J_{ir}(- k \sqrt{192
  \pi} \eta^{-7}/7 a_P M_P)$$ 
  and 
  $$  \phi^{(2)}_k(t) = C_2 Y_{ir}(- k \sqrt{192
  \pi} \eta^{-7}/7 a_P M_P)$$ are normalized. As a result, $C_1$ and $C_2$ are
evaluated as
$$ C_1 = C_2 = \Big[\frac{24 \pi^2 k}{7 a_P M_P^2 \eta_1^7 sin (r \pi)}
\Big]^{1/2} . \eqno(3.26e,f)$$

For large $x$, 
$$J_n(x) \simeq \sqrt{\frac{2}{\pi x}} cos[x - \pi/4 - n \pi/2]  \eqno(3.27a)$$
and
$$Y_n(x) \simeq \sqrt{\frac{2}{\pi x}} sin[x - \pi/4 - n \pi/2]  \eqno(3.27b)$$
Using approximations (3.27a) and (3.27b) in (3.26a), we obtain
$$ \phi_k(\eta) \simeq \sqrt{\frac{7 a_P M_P}{4 \pi \sqrt{3\pi}k}} \eta^{-8}
\Big[ - i C_1 cos (x - \pi/4 - ir \pi/2)$$
$$ + i C_2 sin ((x - \pi/4 - ir \pi/2) \Big], \eqno(3.28a)$$ 
where 
$$ x = \frac{- k \sqrt{192 \pi} \eta^{-7}}{7 a_P  M_P} . \eqno(3.28b)$$
$cosx$ and $sinx$ have similar characteristics as both are periodic
functions. So, there is no harm in taking arbitrary constants $C_1$ and $C_2$
as $C_1 = i C_2.$ With this identification, (3.28a) is obtained as 
$$\phi_k(\eta) \simeq C_2 \sqrt{\frac{7 a_P M_P}{4 \pi \sqrt{3\pi}k}} e^{-
  r\pi/2} \eta^{-8} e^{i ( 8 k \sqrt{3 \pi} \eta^{-7}/7 a_P
  M_P + \pi/4)}  \eqno(3.29a)$$
with
$$ C_2 = \Big[\frac{a_P M_P \eta_1^7}{8 (2\pi)^3 \sqrt{3 \pi}} \Big]^{1/2}. \eqno(3.29b)$$
It is obtained using the normalization condition $(\phi_k, \phi_{k^{\prime}})
= \delta_{kk^{\prime}}$ with scalar product (A6) at the hyper surface $\eta = eta_1.$

Here, $in-$ and $out-$ states are obtained when ${\tilde \eta}$ tends to $- \infty$
and $\infty$ respectively.  So,
\begin{eqnarray*}
 \delta \phi(\eta, {\bf x}) &=& {\sum_{- \infty}^{\infty}} [\phi^{out}_k (\eta) a^{out}_k
e^{i {\vec k}.{\vec x}} + \phi^{ out *}_k (\eta) a_k^{out \dag} 
e^{-i {\vec k}.{\vec x}} ] \\  &=& {\sum_{- \infty}^{\infty}} [\phi^{in}_k (\eta) a^{in}_k
e^{i {\vec k}.{\vec x}} + \phi^{in *}_k (\eta) a_k^{in \dag } 
e^{-i {\vec k}.{\vec x}} ]
\end{eqnarray*}
\vspace{-1.8cm}
\begin{flushright}
(3.30)
\end{flushright}

$\phi^{out}_k (\eta)$ and $\phi^{in}_k (\eta)$ are related through through
Boglubov transformations (A7), where Boglubov coefficients $\alpha_k$ and
$\beta_k$ are defined by (A8) and (A9) respectively and satisfy the condition
(A10).

Using the above definition of $in-$ and $out-$ states, we have
$$ \phi^{out}_k (\eta) = C_2 \sqrt{\frac{7 a_P M_P}{4 \pi \sqrt{3\pi}k}} e^{-
   r\pi/2} \eta^{-8} e^{i (- 8 k \sqrt{3 \pi} \eta^{-7}/7 a_P
  M_P + \pi/4)}  \eqno(3.31a)$$
and
$$ \phi^{in}_k (\eta) = C_2 \sqrt{\frac{7 a_P M_P}{4 \pi \sqrt{3\pi}k}} e^{-
  r\pi/2} \eta^{-8} e^{i ( 8 k \sqrt{3 \pi} \eta^{-7}/7 a_P
  M_P + \pi/4)}  \eqno(3.31b)$$

Using (3.31a) and (3.31b) in (A8) and (A9), Boglubov coefficients are obtained
as
$$ \alpha_k = - 4 i C_3^2 \eta_1^{-16} [8 \eta_1^{-1} - 7iC_4] e^{-2 i C_4
  \eta_1^{-7}} \eqno(3.32a)$$
and
$$ \beta_k = 32 i C_3^2 e^{i \pi/2} \eta_1^{-17}  \eqno(3.32b)$$
where
$$ C_3 = C_2 e^{-r \pi/2}\sqrt{7 a_P M_P/ 4 \pi k \pi} \eqno(3.32c)$$
and
$$ C_4 = 8 k \sqrt{3 \pi}/7 a_P M_P . \eqno(3.32d)$$

Introducing (3.32a) and (3.32b) in conditions (A10), it is obtained that
$$ 28 C_3^2 C_4 \eta_1^{-16} =  1  \eqno(3.33a)$$
yielding 
$$\eta_1 = \Big[\frac{C_2^2 e^{-r \pi}}{\pi} \Big]^{1/16}  \eqno(3.33b)$$
So,
$$|\alpha_k|^{2} = 1 + 1024 C_3^4  \eta_1^{- 34} = 1 + \frac{{\tilde A}}{k^2} \eqno(3.34a)$$ 
with
$$ {\tilde A} = 1024 [C_2 e^{-r \pi/2}\sqrt{7 a_P M_P/ 4 \pi  \pi}]^4
\eta_1^{- 34} . \eqno(3.34b)$$ 

Further non-vanishing $|\beta_k|^2 $, obtained from (3.32b), shows creation of
spinless quantum particles during inflation. 

Moreover, the rate of creation of scalar particle-antiparticle
pairs  per unit time per unit volume \cite{ndb}, due to decay of
$\phi$, is obtained as
\begin{eqnarray*}
 \Gamma_{\phi \to {\bar \Phi}{\Phi}} &=& - ln ({\prod_{K =
 -\infty}^{\infty}}|\alpha_k|^{-2})/V_4 = {\sum_{K =
 -\infty}^{\infty}} ln |\alpha_k|^{2}/V_4 \\ &\simeq & {\sum_{K =
 -\infty}^{\infty}}\frac{{\tilde A}}{k^2}/V_4 = {\tilde A} \zeta(2)/V_4 \\ &=&
 \frac{25 M_P {\tilde A} \zeta(2)}{8 \sqrt{3 \pi} a_P^{24} \eta^{25}}.
\end{eqnarray*}
\vspace{-1.8cm}
\begin{flushright}
(3.35a)
\end{flushright}
where Riemann zeta function $\zeta(2) = \pi^2/6$ and
$$ V_4 = \int_{t_P}^{t_c} a^3 (t) {dt} = \frac{8 \sqrt{3 \pi} a_P^{24}
  \eta^{25}}{25 M_P} .  \eqno(3.35b)$$ 

(3.35a) shows that creation rate of these particles decreases as time increases.

\smallskip

\noindent \underline{\bf Creation of spin-1/2 Dirac fermions}
\smallskip

If $\psi(t, {\bf x})$ are spin-1/2
Dirac spinors (being mathematical representation of spin-1/2 elementary particles) during this
phase , it is natural for these quantum fields to interact with curvature
induced quintessence scalar 
$\phi$ with interaction term $- h {\bar
  \psi} \phi_0 \psi ({\bar \psi} = \psi^{\dag} \gamma^0$, where 
$\psi^{\dag}$ is the adjoint of $\psi$ and $\gamma^0$ is the time-component
of the Dirac matrix in flat space-time. Here $h$ is the dimensionless
coupling constant. In the present model,it is assumed that few spin-1/2
elementary particles exist during this period, but these are not sufficient to
effect cosmic dynamics. Rather, these particles act as seeds for creation of more
particles during the inflationary period. 

From the action  
$$ S_{\psi} = \int {d^4x} \sqrt{-g} {\bar \psi}[i \gamma^{\mu} D_{\mu} -
h \phi_0 ]\psi,  \eqno(3.36a)$$ 
the Dirac equation is obtained as
$$ i \gamma^{\mu} D_{\mu} \psi - h \phi_0 \psi = 0  \eqno(3.36b)$$
where $D_{\mu}$ is defined in (A11).

In the space-time (2.12), (3.36b) is obtained as
$$ [a(t) {\tilde \gamma}^0 \partial_0 + {\tilde \gamma}^a \partial_a  + i h \phi_0 a(t)]
  \psi = 0  \eqno(3.36c)$$
with $ a = 1,2,3.$
Operating (3.36c) by $[a(t) {\tilde \gamma}^0 \partial_0 + {\tilde \gamma}^a
  \partial_a - i h \phi_0 a(t)]$ from left and using the  decompositions of
  $\psi$ given in Appendix A, we get the differential equation for $f_{k,s}
  (g_{k,s})$ as

$$  {\ddot {\tilde f}} + \frac{\dot a}{a}{\dot {\tilde f}} + [\frac{k^2}{a^2}
+ i \epsilon \frac{h}{a}  ({\dot \phi_0} a + {\dot a} \phi_0) + h^2
\phi_0^2 ]{\tilde f} = 0  , \eqno(3.36d)$$ 
where ${\tilde f} = f_{k,s} (g_{k,s}).$

(3.10) shows that for $t < 25.56 t_P$,
$$ \tau = \frac{M_P}{\sqrt{3 \pi}} (t - t_P) < 1 .\eqno(3.37a)$$
So, for $t < 25.56 t_P$, $a(t)$ (given by (3.10)) is approximated to
$$ a(t) = a(\tau) \simeq a_P e^{\tau}  .\eqno(3.37b)$$ 

(3.20) and (3.37b) imply
$$ \phi_0 \simeq  \frac{M_P}{\sqrt{32\pi}} \tau   \eqno(3.37c)$$
for $t < 25.56 t_P$.

Using (3.37a), (3.37b) and (3.37c), (3.36d) is obtained as
$$  \frac{d^2{\tilde f}}{d\tau^2} + \frac{d{\tilde f}}{d\tau} +  \Big[\frac{3 \pi
  k^2}{a_P^2 M_P^2}(1 - 2 \tau + 2 \tau^2) 
+ i \epsilon h \sqrt{\frac{3}{32 }} (1 +  \tau) + \frac{3}{32} h^2 \tau^2\Big]{\tilde f} = 0  , \eqno(3.38)$$ 

(3.38) is obtained in the simplified form as
$$  \frac{d^2{\tilde f}}{d{\tilde \tau}^2} + \frac{d{\tilde f}}{d{\tilde
  \tau}} +  \Big[\frac{K^2}{M^2} + \frac{L^2}{M^2} e^{2 {\tilde \tau}}\Big]{\tilde f} = 0  , \eqno(3.39a)$$
where
$$ K^2 = \frac{3 \pi k^2}{a_P^2 M_P^2} + i \epsilon h - \frac{(- 6 \pi k^2 + i
  \epsilon h a_P^2 M_P^2)}{24 \pi k^2 a_P^2 M_P^2 + 3 h^2 a_P^2 M_P^2/8 }
  ,\eqno(3.39b)$$

$$ L^2 = \frac{(- 6 \pi k^2 + i \epsilon h a_P^2 M_P^2 \sqrt{3/32})}{\sqrt{6 \pi k^2 a_P^2 M_P^2 + 3 h^2 a_P^2 M_P^2/32 }}
  \eqno(3.39c)$$
and
$${\tilde \tau} = M \tau  \eqno(3.39d)$$  
with
$$ M = L^{-2} [- 6 \pi k^2 + i \epsilon h a_P^2 M_P^2 \sqrt{3/32}]. \eqno(3.39e)$$
(3.39a) yields the solution
$$ {\tilde f} = C  e^{- {\tilde \tau}/2} J_{i\sqrt{K^2 - 1/4}/M}(L e^{\tilde
  \tau}/M ) . \eqno(3.40)$$ 

Using (3.40) in (A14) and (A15), it is obtained that
$$\psi_{I k,s} = C  e^{- {\tilde \tau}/2} J_{i\sqrt{K^2 - 1/4}/M}(L e^{\tilde
  \tau}/M )  e^{- i {\vec k}.{\vec x}} { u}_s,  \eqno(3.41a)$$
$$\psi_{II k,s} = C  e^{- {\tilde \tau}/2} J_{i\sqrt{K^2 - 1/4}/M}(L e^{\tilde
  \tau}/M )  e^{i {\vec k}.{\vec x}} {\hat u}_s,  \eqno(3.41b)$$
with ${ u}_s$ and ${\hat u}_s$ given in Appendix A.

Using $in$- and $out$-states defined as above, we obtain $\psi^{in}_{Ik,s}$ and
$\psi^{out}_{I k, s}$ for ${\tilde \tau} < 0$ and ${\tilde \tau} > 0$
respectively. Similarly, $\psi^{in}_{II(-k,-s)}$ and
$\psi^{out}_{II(-k,-s)}$ are obtained.

In this case, the Bogolubov coefficients $\alpha_{k,s}$ and $\beta_{k,s}$,
defined by (A17) and (A18), are obtained as
$$ | \alpha_{k,s}|^2 = \frac{1}{2} = |\beta_{k,s}|^2   \eqno(3.42a,b)$$
as these satisfy the condition (A16).

When $t > 25.56 t_P$, $a(t)$ (given by (3.10)), $\frac{M_P}{\sqrt{3 \pi}} (t -
t_P) > 1$ and $a(t)$ is approximated to
$$ a(t) \simeq a_P \Big[\frac{M_P}{\sqrt{3 \pi}} \Big]^8 (t - t_P)^8
.\eqno(3.43a)$$ 
as well as 
(3.20) and (3.43a) imply
\begin{eqnarray*}
 \phi_0 &\simeq &  8\frac{M_P}{\sqrt{32\pi}}[ ln(t - t_P) + ln (M_P/{\sqrt{3
    \pi}}) ] \\ &\simeq & 8\frac{M_P}{\sqrt{32\pi}}[ - (t - t_P) + 1 + ln
    (M_P/{\sqrt{3 \pi}}) ]     
\end{eqnarray*}
\vspace{-1.8cm}
\begin{flushright}
(3.43b)
\end{flushright}

So, (3.36d) looks like
$$  {\ddot {\tilde f}} + \frac{8}{(t - t_P)}{\dot {\tilde f}} +
\Big[\frac{k^2}{a_P^2} \Big[\frac{\sqrt{3 \pi}}{M_P (t - t_P)} \Big]^{16} $$
$$+ 8i \epsilon h \frac{M_P}{\sqrt{32\pi}(t - t_P)}(9  + 8 ln
    (M_P/{\sqrt{3 \pi}}) - 8 (t - t_P)) $$
    $$+ 2 h^2 \frac{M^2_P}{\pi}\{[(t -
    t_P) - 1 - ln (M_P/{\sqrt{3 \pi}}) \}^2  \Big]{\tilde f} = 0  , \eqno(3.44a)$$ 

Now, for $  25.56 t_P < t \lesssim 7.77 \times 10^{4} t_P$, (3.44) is
approximated to
$$  {\ddot {\tilde f}} + \frac{8}{(t - t_P)}{\dot {\tilde f}} +
\Big[\frac{k^2}{a_P^2} \Big[\frac{\sqrt{3 \pi}}{M_P (t - t_P)} \Big]^{16} 
 + 2 h^2 \frac{M^2_P}{\pi}(t -  t_P)^2  \Big]{\tilde f} = 0  , \eqno(3.44b)$$ 

For $k$ small, (3.44b) is  approximated as  
$$  {\ddot {\tilde f}} + \frac{8}{(t - t_P)}{\dot {\tilde f}} +
2 h^2 \frac{M^2_P}{\pi}(t -  t_P)^2 {\tilde f} = 0  , \eqno(3.44c)$$ 
which integrates to 
$$ {\tilde f} = C (t - t_P)^{-7/2} J_{-7/4} ( \pm h M_P (t- t_P)^2 /\sqrt{2
  \pi}) . \eqno(3.44d)$$

For $k$ large, (3.44b) is  approximated as  
$$  {\ddot {\tilde f}} + \frac{8}{(t - t_P)}{\dot {\tilde f}} +
\frac{k^2}{a_P^2} \Big[\frac{\sqrt{3 \pi}}{M_P (t - t_P)} \Big]^{16}  {\tilde f} = 0  , \eqno(3.44e)$$ 
which integrates to 
$$ {\tilde f} = C (t - t_P)^{-7/2} J_{-1/2} (-7b (t- t_P)^{-7} ) . \eqno(3.44f)$$
Solutions (3.44d) and (3.44f) also yield  Bogolubov coefficients
$\alpha_{k,s}$ and $\beta_{k,s}$ given by (3.42a,b).

Results (3.42a,b) show that elementary spin-1/2 particle-antiparticle pairs are
created during the inflationary prod. The rate of the creation is obtained as
$$\Gamma_{\phi \to {\bar \psi}{\psi}} = = {\sum_{s = \pm 1}}ln ({\prod_{k =
   -\infty}^{\infty}}|\alpha_{k,s}|^{-2}/V_4 =  {\sum_{s = \pm 1}}{\sum_{k =
   -\infty}^{\infty}} ln |\alpha_{k,s}|^{-2}/V_4 $$
$$ = 4 ln 2/V_4 = 100 M_P ln2/ a_P^3 \sqrt{3 \pi} \Big[1 + M_P (t -
t_P)/sqrt{3 \pi} \Big]^25  \eqno(3.45)$$ 
using $\zeta(0) = - 1/2$ obtained through analytic continuation.

\smallskip

\noindent \underline{\bf Fluctuation of $\phi_0(t)$ during inflation  and
  thermalization}  
\smallskip

(3.13b) is an equation for damped
   harmonic oscillator $\phi_0(t)$, which is the curvature-induced quintessence scalar given by
   (3.20). It indicates small oscillation around $\phi_0(t)$ at
   every instant of time during its fall from high hill of the potential to
   the ground state given by $V(\phi_c) = 0.$ For example, at $t = {\tilde
   t}$, the background scalar is $\phi_0({\tilde t})$. So, small homogeneous
   fluctuation $\delta\phi_0(t) $ (it is taken homogeneous being non-quantum
   and due to homogeneous space-time ) around $\phi_0({ t}) = \phi_0({\tilde
   t})$ will satisfy the equation
$$ {\ddot \delta\phi_0(t)} + 3 \frac{\dot a}{a}{\dot \delta\phi_0(t)} +
V^{\prime\prime}(\phi)_{\phi = \phi_0({\tilde t})} \delta \phi_0(t) = 0
\eqno(3.46)$$ 
which is obtained introducing $\phi_0({ t}) = \phi_0({\tilde t}) + \delta\phi_0({ t})$ in (3.13b).

 Further, incorporating (3.10) and (3.22), (3.46) is obtained as
$$ \frac{d^2 \delta\phi_0(t)}{d \eta^2} + \frac{24}{\eta}
  \frac{d \delta\phi_0(t)}{d \eta} + b^2 \delta \phi_0(t) = 0 \eqno(3.47a)$$
  with 
$$ b^2 = 240 \pi \times 10^{37} e^{\phi_0(t) M_P^{-1}  \sqrt{\pi/8}} . \eqno(3.47b)$$

(3.47a) integrates to
$$\delta\phi_0(t) = C \eta^{-23/2} J_{-23/2}(b \eta) \simeq \sqrt{2/\pi
  b}\eta^{-12} cos (b \eta - \pi/2 + 23 \pi/8) .\eqno(3.48)$$

(3.48) shows that fluctuation $\delta\phi_0(t)$ oscillates
around $\phi_0  = \phi_0({\tilde t})$ with amplitude decreasing with time. So,
energy will be released as radiation due to these fluctuation (as (amplitude)$^2$ yields
energy for the classical field).

As this phenomena will occur at every instant of time during the inflationary
period $t_P < t < t_c$, the continuous release of energy during fall of
$\phi_0$ will thermalize the created elementary particles. The energy
released, during the inflationary period, has density
$$ V(0) - V(\phi_c) \simeq 10^{75} {\rm GeV}^4 . \eqno(3.49)$$

It shows that the released energy will thermalize created elementary
particles upto sufficiently high temperature $ T_c $ such that
$$\frac{\pi^{2}}{15} T_c^{4} = 10^{75}.  $$
It gives
$$ T_c = 4.8 \times 10^{18} {\rm GeV}. \eqno(3.50)$$
Like \cite{sks08a}, here also, this radiation is recognized as cosmic microwave
background radiation (CMB). 

Thus, from the above analysis, it is obtained that, during the inflationary
period, curvature-induced quintessence scalar causes creation of elementary
particles and energy released in the form of radiation heats up these created
particles upto very high temperature. So, these particles are highly
relativistic having thermal equilibrium with the emitted radiation. 

\smallskip

\centerline {\underline{\bf 4. Deceleration driven by radiation and matter as
    well as }}

\centerline {\underline{\bf Acceleration in late and future universe}}
\smallskip

\noindent {\underline{\bf Deceleration driven by radiation }}

(3.4) shows that curvature-induced quintessence energy density vanishes at $a
      = a_c$. As a consequence, universe exits from the inflationary
      phase. During this period, dark energy causes elementary particle -
      antiparticles as well as CMB thermalizing the created particles upto
      temperature $ T_c = 4.8 \times 10^{18} {\rm GeV}$ given by (3.50).

Moreover, in (3.3), we get curvature-induced dark radiation term. Thus, we
have two sources of radiation. As temperature of CMB, obtained here, is very
high, dark radiation too will have thermal equilibrium with CMB. So, energy
density of created particles, dark radiation and CMB together will have energy
density 
$$ \rho_r = 10^{75} \Big(\frac{a_c}{a} \Big)^4. \eqno(4.1)$$

Thus, at the end of inflation ($a = a_c$), we recover standard model of cosmology
and (3.3) 
reduces to
$$ \Big(\frac{\dot a}{a} \Big)^2 \simeq \frac{8 \pi M_P^2}{30}\Big(\frac{a_c}{a} \Big)^4. \eqno(4.2)$$
 
(4.2) integrates to
$$ a(t) = a_c [1 + 4 M_P \sqrt{ \pi/15 a_c^4} (t - t_c) ]^{1/2}. \eqno(4.3)$$

\noindent {\underline{\bf Matter-dominance and deceleration}}

 In the late universe, the effective
Friedmann equation is given by (2.21). Using (2.23)-(2.25) in (2.21), we
obtain 
$$ \Big(\frac{\dot a}{a}\Big)^2 = \frac{C}{a^3} + \frac{8 D}{15} a^{3/4}
\Big[1 - \frac{5 E}{6 D} a^{3/4} \Big] , \eqno(4.4)$$
where
$$ D = \frac{a_c^{-3/4}}{4 \pi G}\Big[1 - \frac{28 \pi G}{135}
\Big(\frac{1}{180 \beta} \Big)^{1/3} a_c^{-1/4} \Big] \eqno(4.5a)$$ 
and
$$ E = \frac{a_c^{-3/2}}{4 \pi G}\Big[1 - \frac{4 \pi G}{135}
\Big(\frac{1}{180 \beta} \Big)^{1/3} a_c^{-1/4} \Big] \eqno(4.5b)$$ 
being obtained from (2.20a) and (2.20b) using (2.23)-(2.25) and (3.2b).

The first term, on r.h.s. of (4.4), emerges spontaneously and has the form of
matter density, so it is recognized as dark matter density like dark
radiation. 

Before proceeding further, it is useful to remark that, like radiation, we
have two types of matter too (i) dark matter given as $\frac{ 3 C}{8 \pi G
  a^3}$ in (4.4a), which is non-baryonic and (ii) baryonic matter formed by
elementary particles ( produced during inflation) through various
processes of 
standard cosmology such as nucleosynthesis, baryosynthesis and recombination
of hydrogen (not being discussed here).

According to WMAP results \cite{abl}, present density of pressureless 
matter (baryonic and non-baryonic) is obtained to be $\rho^{(m)}_0 = 0.27\rho_0^{\rm cr}$ and present dark
energy density $\rho^{\rm ph}_{{\rm de}0} = 0.73\rho_0^{\rm cr}$      with
$$\rho_0^{\rm cr} = \frac{3 H_0^2}{8 \pi G},$$
where current Hubble's rate of expansion $H_0 = 100h km/Mpc second = 2.32
\times 10^{-42} h {\rm GeV}$ and $h = 0.68$. Thus,
$$ \rho_0^{\rm cr} = 2.9 \times 10^{-47} {\rm GeV}^4. \eqno(4.10a)$$
$$H_0  = 1.58 
\times 10^{-42} h {\rm GeV} = [0.96 t_0]^{-1} \eqno(4.10b)$$
and
$$t_0 = 13.7 {\rm Gyr} = 6.6 \times 10^{41} {\rm GeV}^{-1}. \eqno(4.10c)$$

Using these values, it is obtained that
$$ \rho^{(mat)} = \frac{ 3 C}{8 \pi G a^3} = \frac{0.27 H^2_0}{a^3}
\eqno(4.11a)$$
with the present scale factor normalizing as
$$ a_0 = 1 \eqno(4.11b)$$
and
$$ \rho^{\rm ph}_{\rm de} = \frac{D}{5 \pi G} a^{3/4} = {0.73 H_0^2}{a^{3/4}}  \eqno(4.12)$$
from (4.6).

Connecting (4.4), (4.7) and (4.8), it is obtained that
$$ \Big(\frac{\dot a}{a}\Big)^2 =  \frac{0.27 H^2_0}{a^3} + 0.73 H_0^2 a^{3/4}
\Big\{1 - \frac{0.73 \rho_0^{\rm cr} a^{3/4}}{2 \lambda} \Big\} \eqno(4.13)$$

(4.13) shows that
$$\frac{0.27 H^2_0}{a^3} > 0.73 H^2_0 a^{3/4}$$   
for $ a < 0.767$ and 
$$\frac{0.27 H^2_0}{a^3} < 0.73 H^2_0 a^{3/4}$$   
for $ a > 0.767.$

It means that a {\em transition} from matter-dominance to DE-dominance takes place at
$$ a_* = 0.767\eqno(4.14)$$
giving {\em red-shift}
$$ z_* = \frac{1}{a_*} - 1 = 0.303 \eqno(4.15)$$
which is very closed to lower limit of $z_*$ given by 16 Type supernova observations \cite{ag}.
Thus, for $ a < 0.767$, (4.13) is approximated as
$$ \Big(\frac{\dot a}{a}\Big)^2 = \frac{0.27 H^2_0}{a^3}  \eqno(4.16)$$
using matter density given by (4.11a) and 
$$[{0.73 \rho_0^{\rm cr} a^{3/4}}]^2 << {0.73 \rho_0^{\rm cr} a^{3/4}}$$ 
for $a $ being less than $1$.
 
(4.16) integrates to
$$a(t) = a_d [1 + \frac{3}{2} \sqrt{0.27} H_0 a_d^{-3/2}(t - t_d)]^{2/3}
. \eqno(4.17)$$

(4.13) is approximated as (4.16) when
$$ \frac{0.27 H^2_0}{a^3} >>  0.73 H_0^2 a^{3/4},$$
but as $a(t)$ approaches very close to $a = a_*$, 
$$ \frac{0.27 H^2_0}{a^3} \approx  0.73 H_0^2 a^{3/4}$$
So, (4.16) needs to be modified as
$$ \Big(\frac{\dot a}{a}\Big)^2 = \frac{0.54 H^2_0}{a^3} ,  \eqno(4.18a)$$
which integrates to
$$a(t) = a_d [1 + \frac{3}{2} \sqrt{0.54} H_0 a_d^{-3/2}(t - t_d)]^{2/3}
. \eqno(4.18b)$$

It shows decelerated expansion as ${\ddot a} < 0.$ In (4.17), $t_d = = 386
{\rm kyr} = 2.8 \times 10^{-5} t_0 $ (WMAP result) is the decoupling time and
the scale factor $a_d$ at $t = t_d$ is given by $1/a_d = 1 + z_d = 1090$ with 
$z_d = 1089$ (WMAP result). Using these values and $a_*$ from (4.14), (4.10b)
and (4.17)
yields
$$ t_* = 0.63 t_0 . \eqno(4.18c)$$
\smallskip

\noindent {\underline{\bf Phantom dominance and late acceleration}}

When $ a \ge 0.735$, (4.13) is approximated as
$$\Big(\frac{\dot a}{a} \Big)^2 = 0.73 H_0^2 a^{3/2} \Big[ a^{-3/4} -
  \frac{0.73 \rho^0_{\rm cr} }{2 \lambda} \Big]    \eqno(4.19)$$
with $H_0$ given by (4.10b).  

(4.19) integrates to
\begin{eqnarray*}
 a(t) &=&  \Big[ \frac{0.73 \rho^0_{\rm cr}} {2 \lambda} + 
\Big\{\sqrt{ 1.22 - \frac{0.73 \rho^0_{\rm cr}} {2 \lambda}}
\\ &&
- \frac{3}{8} H_0 \sqrt{0.73} (t - t_*)
\Big\}^2 \Big]^{- 4/3}
\end{eqnarray*}
\vspace{-1.7cm}
\begin{flushright} 
(4.20a)
\end{flushright}
as $ a_*^{-3/4} = 1.22$. (4.20a) shows and acceleration and it is singularity-free.

From (4.20a), it is obtained that
$$ {\ddot a} = 0.27 H_0^2 a^{5/2} \Big[ \frac{1.7 \rho^0_{\rm cr}}{\lambda} - \frac{11}{3}a^{-3/4} \Big]. \eqno(4.20b)$$
This shows  ${\ddot a} > 0$ , when 
$$\frac{1.7 \rho^0_{\rm cr}}{\lambda} > \frac{11}{3}a^{-3/4}. $$

Thus, solution (4.20a) yields acceleration in the late universe.

(4.19) shows that accelerated expansion (4.20a) stops at $a = a_e$ satisfying
     the  condition
$$0.73 \rho^0_{\rm cr} a_e^{3/4} = 2 \lambda . \eqno(4.21)$$

Further, (4.20a) yields
 \begin{eqnarray*}
 1 = a_0 &=&  \Big[ \frac{0.73 \rho^0_{\rm cr}} {2 \lambda} + 
\Big\{\sqrt{ 1.22 - \frac{0.73 \rho^0_{\rm cr}} {2 \lambda}}
\\ &&
- \frac{3}{8} H_0 \sqrt{0.73} (t_0 - t_*)
\Big\}^2 \Big]^{- 4/3}.
\end{eqnarray*}
\vspace{-1.7cm}
\begin{flushright} 
(4.22)
\end{flushright}

Using (4.20a) and (4.10b) in (4.22), $\lambda$ is evaluated as
$$ \lambda = 1.17 \rho^0_{\rm cr} .  \eqno(4.23)$$

$a(t)$, given by (4.20a), acquires the value $a_e$ by the time
\begin{eqnarray*}
t_e &=& t_*  + \frac{8}{3} [H_0 \sqrt{0.73}]^{-1}\sqrt{ 1.22 - \frac{0.73
    \rho^0_{\rm cr}} {2 \lambda}} \\&=&  0.63 t_0   + 2.49 t_0 = 3.12 t_0
\end{eqnarray*}
\vspace{-1.7cm}
\begin{flushright} 
(4.24)
\end{flushright} 

\smallskip

\centerline {\underline{\bf 5. Re-dominance of matter, collapse in the future}}

\centerline {\underline{\bf    universe and its avoidance}}

\bigskip
\smallskip

\noindent {\underline{\bf (a)Re-dominance of matter and deceleration in future universe}}

It is shown above that phantom-dominance ends at $t = t_e$. As a consequence, matter re-dominates and (4.13) reduces to
\begin{eqnarray*}
 \Big(\frac{\dot a}{a}\Big)^2 &=&  \frac{0.27 H^2_0}{a^3} + 0.73 H_0^2 a^{3/4}
\Big\{1 - \frac{0.73 \rho_0^{\rm cr} a^{3/4}}{2 \lambda} \Big\}\\ &
\simeq & \frac{0.27 H^2_0}{a^3}
\end{eqnarray*}
\vspace{-1.7cm}
\begin{flushright} 
(5.1)
\end{flushright} 

(5.1) integrates to
$$a(t) = a_e [1 + \frac{3}{2} ]\sqrt{0.27} H_0 a_e^{-3/2}(t - t_e)]^{2/3}
 \eqno(5.2)$$ 
 showing decelerated expansion as ${\ddot a} < 0.$

Moreover, though at $a = a_e$
$$ 0.73 H_0^2 a^{3/4}
\Big\{1 - \frac{0.73 \rho_0^{\rm cr} a^{3/4}}{2 \lambda} \Big\} $$
vanishes, it will be negative for $a > a_e$. As a result, at a certain value
of $a(t)$ being $a_m > a_e$, ${\dot a} = 0$. Thus $a_m $ is the maximum of
$a(t)$. It is because, for $a > a_m$, $[{\dot a}/a]^2 < 0$ leading to
imaginary expansion rate ${\dot a}/a.$ So, $a_m $ satisfies the condition
$$  a_m^{15/4}
\Big\{\frac{0.73 \rho_0^{\rm cr} a_m^{3/4}}{2 \lambda} - 1\Big\} = \frac{27 }{73}. \eqno(5.3a)$$ 

(5.3a) yields 
$$ a_m = 0.86/[\rho_0^{\rm cr}]^{2/9} = 1.88 \times 10^{10} \eqno(5.3b)$$ 
 with $ \rho_0^{\rm cr}$ given by (4.10a)

Connecting (4.21), (4.23), (4.24), (5.3a) and (5.3b), it is obtained that
$$ t_m \simeq 3,45 \times 10^{15} t_0 .  \eqno(5.3c)$$

\smallskip

\noindent {\underline{\bf (b) Contraction in the future universe and its
    collapse}}

Further, it is interesting to note that the curve $a = a(t)$ will be
continuous at $t = t_m$, but the direction of tangent to this curve (pointed
at $a = a_m$) will
change yielding ${\dot a} < 0$ for $t > t_m$. It means that
universe will retrace back at $t = t_m$ and will begin to contract for $t >
t_m$.  During the contraction phase, $a(t)$ will decrease with time. So, term proportional to $a^{-3}$ will dominate over terms
proportional to $a^{3/4}$ in (4.13). As a result, we have
$$ H^2 = \Big(\frac{\dot a}{a} \Big)^2  \simeq {0.27} H_0^2 a^{-3} . \eqno(5.4a)$$
yielding
$$ H = \Big(\frac{\dot a}{a} \Big) \simeq - \sqrt{0.27} H_0 a^{-3/2} . \eqno(5.4b)$$
(5.4b) integrates to 
$$a(t) = a_m [1 - \frac{3}{2} \sqrt{0.27} H_0 a_m^{-3/2}(t - t_m)]^{2/3}
 \eqno(5.5)$$ 
 showing decelerated contraction  as ${\ddot a} < 0.$

(5.5) yields $a(t) = 0$ at
\begin{eqnarray*} 
 t = t_{\rm col} &=& t_m +  \frac{2}{3\sqrt{0.27}} H_0^{-1} a_m^{3/2}\\& =& 3.62\times 10^{15} t_0 .
\end{eqnarray*}
\vspace{-1.7cm}
\begin{flushright} 
(5.6)
\end{flushright}

So, at $ t = t_{\rm col}$, dominating energy density term 
$$ \rho^{(mat)} = 0.27 \rho_0^{\rm cr}/ a^3,  \eqno(5.7)$$ 
in (5.1), will be infinite. These results show cosmic collapse at this particular epoch.

In what follows, a possibility to avoid the cosmic collapse, obtained by the
classical mechanics, is demonstrated.

\smallskip

\noindent {\underline{\bf (c) Scalar field for matter and particle}}
\noindent {\underline{\bf creation in the
    future}}
    
\noindent {\underline{\bf     universe}}

In section 3(b), curvature-induced quintessence dark energy is realized
through the scalar $\phi$. Using the same approach, matter term (5.7), in
Friedmann equation (5.1), can be realized through a different scalar
$\Phi(x,t)$ obeying the equation
$$ {\Box} \Phi(t,{\bf x}) +  V^{\prime}(\Phi) = 0. \eqno(5.8a)$$
 Here, $V^{\prime}(\Phi) = d V/d\Phi$ and $\Phi(x,t)$ has mass dimension equal to 1. 

In the homogeneous and flat model of the universe
(2.12), (5.8a) is obtained as
$${\ddot \Phi(t)} + 3 \frac{\dot a}{a}{\dot \Phi(t)} + V^{\prime}(\Phi)_{\Phi = \Phi(t)} = 0
\eqno(5.8b)$$
as $\Phi(t,{\bf x}) = \Phi(t)$ due to homogeneity. 

Density and pressure for $\Phi(t)$ are given as
$$ \rho^{(mat)} = \frac{1}{2} {\dot \Phi}^2 + V(\Phi) \eqno(5.9a)$$
and
$$ 0 = p^{(mat)} = \frac{1}{2} {\dot \Phi}^2 - V(\Phi) \eqno(5.9b)$$
 
Connecting (5.4), (5.7), (5.9a) and (5.9b). it is obtained that
$$\frac{\dot a}{a}  \simeq - \sqrt{8 \pi/3} M_P^{-1} {\dot \Phi}
. \eqno(5.10a)$$

(5.10a) integrates to
$$ a(t) = a_m e^{- \sqrt{8 \pi/3} M_P^{-1} (\Phi - \Phi_{ m})}
. \eqno(5.10b)$$ 

(5.7), (5.9a) and (5.9b) yield
\begin{eqnarray*}
 V(\Phi) &=&  0.135 \frac{\rho_0^{\rm cr}}{a^3} \\&=& 0.135 \frac{\rho_0^{\rm cr}}{a_m^3}e^{ \sqrt{24 \pi} M_P^{-1}
  (\Phi - \Phi_{ m})} .
\end{eqnarray*}
\vspace{-1.7cm}
\begin{flushright} 
(5.10c)
\end{flushright}

It is discussed above that near $t = t_{\rm col}$, energy density will be
extremely high. This situation is analogous to the state of early
universe with high energy density and large curvature. So, near collapse time,
classical mechanics is not the appropriate theoretical machinery and
investigations should be done using quantum field theory. These arguments
prompt us to resort to creation of quantum particles due to gravitational
changes and probe its
back-reaction on the future universe.

If $\delta\Phi(t,{\bf x})$ is the quantum fluctuation in homogeneous background
field $\Phi(t)$, we have the equation of motion for $\delta\Phi(t,{\bf x})$ as
 $$ {\Box} \delta\Phi(t,{\bf x}) +  V^{\prime\prime}(\Phi)|_{\Phi = \Phi(t)}\delta\Phi(t,{\bf x})  = 0 \eqno(5.11)$$
obtained from (5.8a). 

In the space-time (2.12),(5.11) is re-written as
$$ {\ddot \delta\Phi(t,{\bf x})} + 3 \frac{\dot a}{a}{\dot \delta\Phi(t,{\bf x})} - a^{-2} \Big[
\frac{\partial^2}{\partial x^2} + \frac{\partial^2}{\partial y^2} +
\frac{\partial^2}{\partial z^2} \Big]\delta\Phi (t, {\bf x})$$
$$  +
V^{\prime\prime}(\Phi)_{\Phi = \Phi(t)} \delta \Phi(t,{\bf x}) = 0
.\eqno(5.12)$$

Using the decomposition
$$  \delta \Phi(t, {\bf x}) = {\sum_{- \infty}^{\infty}} [\Phi_k (t) a_k
e^{i {\vec k}.{\vec x}} + \Phi^*_k (t) a_k^{\dag}
e^{-i {\vec k}.{\vec x}} ] \eqno(5.13)$$
as well as results (5.10c) and $a(t)$ given by (5.5), (5.12) is obtained as
$$ \Phi_k^{\prime\prime}({\tilde \eta}) + \frac{2}{{\tilde \eta}} \Phi_k^{\prime\prime}({\tilde \eta}) 
+ \Big\{ \Big(\frac{2 k a_m^{1/2}}{3 H_0 \sqrt{0.23}} {\tilde \eta}^{-2/3}\Big)^2  + \frac{2}{{\tilde \eta}^2} \Big\}\Phi_k({\tilde \eta}) = 0  \eqno(5.14a)$$
with
$$ {\tilde \eta} = [1 - \frac{3}{2} \sqrt{0.27} H_0 a_m^{-3/2}(t - t_m)]
. \eqno(5.14b)$$
Here prime ($^{\prime}$) signifies derivative with respect to ${\tilde \eta}$.

(5.14a) integrates to 
$$\Phi_k({\tilde \eta}) = {\tilde \eta}^{-1/2} [C_1 J_in (b {\tilde
  \eta}^{1/3})  + Y_in (b {\tilde   \eta}^{1/3})],  \eqno(5.15a)$$
where
$$ b = \frac{2 k a_m^{1/2}}{ H_0 \sqrt{0.23}} \eqno(5.15b)$$
and
$$ n = 3 \sqrt{7}/2 . \eqno(5.15c)$$

Using identities (3.26c), (3.26d) and (A6), $C_1$ and $C_2$ in (5.15a) are
evaluated as
$$C_1 = C_2 = \Big[\frac{\pi b a_m^{-3/2}}{3 \sqrt{0.27}H_0 sinh(n \pi) V
  {\tilde \eta}_1^{5/3}} \Big]^{1/2} , \eqno(5.15d,e)$$ 
where $V$ is the 3-volume.

Using approximations (3.27a) and (3.27b) for Bessel functions, (5.15a) is
obtained as
$$\Phi_k({\tilde \eta}) \simeq C {\tilde \eta}^{-2/3}e^{-i \pi/4} e^{ i b {\tilde \eta}^{1/3}}  \eqno(5.16a)$$
with 
$$ C =  \Big[\frac{2 \pi a_m^{-3/2} e^{n \pi/2}}{3 \sqrt{0.27}H_0 sinh(n \pi) V
  {\tilde \eta}_1^{5/3}} \Big]^{1/2} .
\eqno(5.16b)$$ 

Like section 3(b), here also $in-$ and $out-$ states are defined such that
$$\Phi^{in}_k({\tilde \eta}) \simeq C {\tilde \eta}^{-2/3}
  e^{-i \pi/4} e^{ i b
  {\tilde \eta}^{1/3}}  \eqno(5.17a)$$
and
$$\Phi^{out}_k({\tilde \eta}) \simeq C{\tilde \eta}^{-2/3}
  e^{-i \pi/4} e^{- i b
  {\tilde \eta}^{1/3}}.  \eqno(5.17b)$$

Using (5.17a) and (5.17b) in (A9), Boglubov coefficient $ |\beta_k|$ is
obtained as
$$ \beta_k = 4 \pi a_m^{3/2} e^{3 \pi \sqrt{7}/4} e^{-i \pi/2} {\tilde
  \eta}_1^{- 2}/3 sinh (3 \pi \sqrt{7}/2) .\eqno(5.18)$$

Further, (5.18) yields
$$ |\beta_k|^2 = 16 \pi^2 a_m^{3} e^{3 \pi \sqrt{7}/2} {\tilde
  \eta}_1^{- 4}/9  sinh^2 (3 \pi \sqrt{7}/2) , \eqno(5.19)$$ 
which is non-vanishing. It shows creation of spinless
particle-antiparticle pairs.

Energy density for the quantum fluctuation $\delta\Phi(t,{\bf x})$, satisfying
(5.12) is obtained as
\begin{eqnarray*}
\rho_{\delta\Phi(t,{\bf x})} &=& \frac{1}{2} {\dot \delta\Phi(t,{\bf x})} {\dot {\delta\Phi(t,{\bf x})}^*} + \frac{1.215 H_0^2}{a^3} {\delta\Phi(t,{\bf
  x})}{\delta\Phi(t,{\bf x})}^*  \\&&
+  \frac{1}{2 a^(t)} \Big[\Big(\frac{\partial \delta\Phi(t,{\bf
  x})}{\partial x}\Big) \Big(\frac{\partial \delta\Phi^*(t,{\bf
  x})}{\partial x}\Big) + \Big(\frac{\partial \delta\Phi(t,{\bf
  x})}{\partial y}\Big) \Big(\frac{\partial \delta\Phi^*(t,{\bf
  x})}{\partial y}\Big) \\&& + \Big(\frac{\partial \delta\Phi(t,{\bf
  x})}{\partial z}\Big) \Big(\frac{\partial \delta\Phi^*(t,{\bf
  x})}{\partial z}\Big)  \Big] . 
\end{eqnarray*}
\vspace{-2cm}
\begin{flushright} 
(5.20)
\end{flushright}
\vspace{0.5cm}

Incorporating (5.13), (5.20) yields
$$\rho_{\delta\Phi(t,{\bf x})} = {\sum_{- \infty}^{\infty}} \Big[  \frac{1}{2}
{\dot \delta\Phi_k(t)} {\dot \delta\Phi^*_k(t)} + \Big(\frac{1.215 H_0^2}{a^3}
- \frac{k^2}{2 a^(t)} + \frac{k^2}{2 a^2} \Big) \delta\Phi_k(t)
\delta\Phi^*_k(t) \Big] . \eqno(5.21)$$

Now, renormalized energy density of created spinless bosons is obtained as
\begin{eqnarray*}
 \rho_{\rm created} &=& < in| \rho^{in} | in > - < in | \rho^{out} | in > \\
 &=& - {\sum_{- \infty}^{\infty}} 0,27 C^2 |\beta_k|^2 H^2_0 [ \frac{4}{9}
 a_m^2 a^{-5} +  9 a_m^2 a^{-5}] \\&& - {\sum_{- \infty}^{\infty}} k^2 |\beta_k|^2 [\frac{1}{2
 a^2} + 0.73 a_m^4 a^{-6} ] \\  &=& - 2 \zeta(0)  0,27 C^2 |\beta_k|^2 H^2_0 [ \frac{4}{9}
 a_m^2 a^{-5} +  9 a_m^2 a^{-5}] \\&&  - 2 \zeta(-2) |\beta_k|^2 [\frac{1}{2
 a^2} + 0.73 a_m^4 a^{-6} ] \\  &=& \frac{680}{27} \sqrt{0.27} \pi^3 a_m^{7/2}
 \Big(\frac{H_0}{V} \Big) \Big(e^{3 \pi \sqrt{7}} sinh^3(3 \pi \sqrt{7}/2)
 {\tilde   \eta}_1^{- 17/3} a^{-5}
 \end{eqnarray*}
\vspace{-3.5cm}
\begin{flushright} 
(5.22)
\end{flushright}
\vspace{2cm}   
using (5.17a), (5.17b) and (5.19), $\zeta(0) = -1/2$ (obtained through
analytical continuation) and $\zeta(-2) = 0.$

\smallskip

\noindent {\underline{\bf (d)Back - reaction of created particles and escape}}

\noindent {\underline{\bf from the
    cosmic collapse}}
    
\smallskip

It is natural to think that created spinless  particles (obtained above) will effect cosmic
dynamics. As a consequence, 
FE (5.4a) is modified as
$$ H^2  \simeq {0.27} H_0^2 a^{-3} + \frac{8 \pi}{3} M_P^{-2} X a^{-5},  \eqno(5.23a)$$ 
when energy density of created particles will be significant. Here,
$$ X = \Big[\frac{680}{27} \sqrt{0.27} \pi^3 a_m^{7/2}
 \Big(\frac{H_0}{V} \Big) \Big(e^{3 \pi \sqrt{7}} sinh^3(3 \pi \sqrt{7}/2)
 {\tilde   \eta}_1^{- 17/3}  \Big]   \eqno(5.23b)$$ 

The solution of (5.23a) can be taken as
$$ a = a_{\rm col} exp [|D (t_{\rm col} - t)\}| + \gamma |D (t_{\rm
  col} - t)|^2],  \eqno(5.24)$$
where $a_{\rm col}$ is the scale-factor at $t = t_{\rm col}$, $D$ is a
  constant of mass dimension and $\gamma$ is a dimensionless constant.

(5.24) yields 
$$ H = - D [ 1 + 2 \gamma |D (t_{\rm   col} - t)|].  \eqno(5.25)$$

(5.24), being solution of (5.23), will satisfy (5.23). So,
$$\frac{3}{8 \pi} M_P^2 D^2 [1 + 4 \gamma |D (t_{\rm   col} - t)| + 4 \gamma^2
|D (t_{\rm   col} - t)|^2 ] \simeq \frac{0.81}{8 \pi} M_P^2 H_0^2 a_{\rm
  col}^{-3}$$
$$\times[
1 - 3 |D (t_{\rm col} - t)| - 3 \gamma |D (t_{\rm   col} - t)|^2]  $$
$$ + X  a_{\rm   col}^{-5} [1 - 5 |D (t_{\rm col} - t)|
 - 5 \gamma |D (t_{\rm   col} - t)|^2]  . \eqno(5.26)$$

(5.26) yields
\begin{eqnarray*}
\frac{3}{8 \pi} M_P^2 D^2 &=& \frac{0.81}{8 \pi} M_P^2 H_0^2 a_{\rm col}^{-3}
+ X a_{\rm   col}^{-5} \\ \frac{3}{2 \pi} M_P^2 D^2 \gamma
 &=& - 3 \frac{0.81}{8 \pi} M_P^2 H_0^2 a_{\rm col}^{-3} - 5 X a_{\rm   col}^{-5} \\ \frac{3}{2 \pi} M_P^2 D^2
 \gamma^2  
 &=& \Big[- 3 \gamma + \frac{9}{2} \Big] \frac{0.81}{8 \pi} M_P^2 H_0^2 a_{\rm
   col}^{-3} + \Big[- 5 \gamma + \frac{25}{2} \Big] X a_{\rm   col}^{-5}  
 \end{eqnarray*}
$$\eqno(5.27a,b,c)$$

comparing constants terms on both sides as well as coefficients of $|D (t_{\rm
  col} - t)|$ and $|D (t_{\rm col} - t)|^2$.

From (5.27b,c), it is obtained that
$$ 25 X a_{\rm   col}^{-5} = - 9 \frac{0.81}{8 \pi} M_P^2 H_0^2 a_{\rm
   col}^{-3}. \eqno(5.28)$$
This equation shows that as universe contracts matter density will decrease and the effect of
    back-reaction of created particles will increase after a certain
   epoch  $t = t_c$.

Using (5.28) in (5.27a,b), it is obtained that
\begin{eqnarray*}
D^2 &=& \frac{4.32}{25}  H_0^2 a_{\rm col}^{-3}
 \\ 4 D^2 \gamma
 &=& -  \frac{1.62}{5}  H_0^2 a_{\rm col}^{-3} .
\end{eqnarray*}
\vspace{-2cm}
\begin{flushright} 
(5.29a,b)
\end{flushright}
\vspace{0.5cm} 

(5.29a) and (5.29b) yield
$$ \gamma  = - \frac{15}{32} . \eqno(5.30)$$
Also (5.29a) shows that $a_{\rm col}$ can not vanish, because $D$ will be divergent. Thus, it is found that back-reaction of created particles
will make energy density finite at $t = t_{\rm col}$.

Planck scale is the fundamental scale. It suggests the largest energy mass
scale as Planck mass $M_P$. At this  scale, energy density is obtained
$M_P^4/8 \pi^2$. So, we have
$$\frac{3}{8 \pi} M_P^2 D^2 = \frac{4.32}{200 \pi} M_P^2 H_0^2 a_{\rm
  col}^{-3} = \frac{M_P^4}{8 \pi^2}  \eqno(5.31)$$
connecting (5.27a) and (5.28).

(5.31) yields
$$ a_{\rm col} = 2.25 \times 10^{-42}  \eqno(5.32)$$
and
$$ D = \frac{M_P}{\sqrt{3 \pi}} .  \eqno(5.33)$$

\smallskip

\centerline {\underline{\bf 6. Salient features and concluding remarks}}

\smallskip
Here, a cosmological picture is obtained from the gravitational action containing the linear Einstein term as well as non-linear terms $ R^2 $ and $ R^5 $. This is a gravitational action with no other field except the scalar curvature $R$. The approach of this paper is different from the approach in Refs.\cite{snj,lds, sn08}. This approach has an advantage to have power to explain 

\noindent (i) power-law inflation in the early universe and graceful exit from this phase, 

\noindent (ii) creation of SM particles, 

\noindent (iii) recovery of the standard cosmology with the cosmic background radiation with extremely high initial temperature $ \sim 10^{18}{\rm GeV}$, 

\noindent (iv) deceleration of the universe driven by radiation emitted during the inflationary phase and particles in thermal equilibrium with radiation, 

\noindent (v) deceleration driven by curvature induced dark matter and baryonic matter caused by various processes like nucleosynthesis, baryosynthesis, hydrogen re-combination of elementary particles created during inflation,

\noindent (vi) dominance of curvature-induced phantom at red-shift $z = 0.303$ (which is consistent with observational results),

\noindent (vii) transient acceleration driven by phantom in the very late universe, 

\noindent (viii) re-dominance of matter, contraction of the universe, collapse of the universe at time $t_{\rm col}= 3.62\times 10^{15} t_0 $ , 

\noindent (ix) avoidance of collapse due to creation of particles near the time $t_{\rm col}$ as well as its back-reaction and

 \noindent (x) rebirth of the universe after $t_{\rm col}$.

Thus model gives the revival of the state of the early universe at time $t_{\rm col}= 3.62\times 10^{15} t_0 $ . Interestingly, above investigations show that the Ricci scalar $R$ has dual role as a geometrical field as well as a physical as physical and geometrical terms in the above theory are obtained from the scalar curvature $R$ noted in earlier works \cite{skp}.

\bigskip
\centerline{\bf Appendix A}

\smallskip

\noindent \underline{\bf Boson case}

In the case of minimal coupling of $\Phi$ to gravity and having interaction
with $\phi$ as $-(1/2) g \phi^2 \Phi^* \Phi$, we have the equation

$$ {\Box} \Phi (t, {\bf x}) + [g \phi^2(t)  + m_{\Phi}^2] \Phi (t, {\bf x}) =
0 \eqno(A1)$$
with $m_{\Phi}$ being mass of $\Phi (t, {\bf x})$.

$\Phi$ is decomposed as
$$ \Phi (t, {\bf x}) = {\sum_{- \infty}^{\infty}} [\Phi_k (t) a_k
e^{i {\vec k}.{\vec x}} + \Phi^*_k (t) a_k^{\dag}
e^{-i {\vec k}.{\vec x}} ]. \eqno(A2)$$

In (A2), $a_k (a_k^{\dag})$ are  creation (annihilation) operators
satisfying quantum conditions
$$ [a_k , a_{k^{\prime}} ] = 0 = [a_k^{\dag} , a_{k^{\prime}}^{\dag} ]$$
$$ [a_k , a_{k^{\prime}}^{\dag} ] = (2 \pi)^3 \delta_{kk^{\prime}}
\eqno(A3,4,5) $$

The scalar product \cite{ndb} is defined as
$$ (\Phi_1, \Phi_2) = - i \int_{t = constant} \sqrt{-g} {d^3x}[ \Phi_1
{\partial}_t \Phi_2^* - \Phi_2^* {\partial}_t \Phi_1  ] \eqno(A6)$$
with $\partial_t$ denoting derivative with respect to time $t$ . 

The $in-$ and $out-$ states of $\Phi$ are obtained at two extremes of the
space-time where space-time is asymptotically Minkowskian. These are connected
through Boglubov transformations
$$ \Phi^{out}_k(t,x) = \alpha_k \Phi^{in}_k(t,x) + \beta_k \Phi^{in *}_k(t,x)
\eqno(A7)$$

This scalar product (A6) shows that $(\Phi^{in}_k, \Phi^{in *}_k) = 0$,
$(\Phi^{in }_k, \Phi^{in }_k) = 1 = (\Phi^{in *}_k, \Phi^{in
  *}_k)$. So 
From (A7), it is obtained that 
$$\alpha_k = (\Phi^{out}_k(t), \Phi^{in}_k(t)), \beta_k = (\Phi^{out}_k(t),
\Phi^{in {*}}_k(t)) \eqno(A8,9)$$
obeying the condition

$$|\alpha_k|^2 - |\beta|^2 = 1 .  \eqno(A10)$$

\smallskip

\noindent \underline{\bf Case of spin-1/2 Dirac fermion}

The equation of Dirac field $\psi$ with mass $m_{\psi}$ is given as
$$ [i \gamma^{\mu} D_{\mu} -  m_{\psi}]\psi = 0  \eqno(A11)$$
with $i = \sqrt{-1}, D_{\mu} = \partial_{\mu} + \frac{1}{4} 
\Gamma^{\rho}_{\sigma\mu}h^{\sigma}_a)g_{\nu\rho}h^{\nu}_b {\tilde \gamma}^a
{\tilde \gamma}^.$ Here $\gamma^{\mu} = h^{\mu}_a{\tilde \gamma}^a (a =
0,1,2,3)$, $g^{\mu\nu} = h^{\mu}_a h^{\nu}_b \eta^{ab},$ 
$\{\gamma^{\mu}, \gamma^{\nu}\} = 2 g^{\mu\nu} $ and $\{{\tilde \gamma}^a,
{\tilde \gamma}^b = 2\eta^{ab}$ with $\gamma^{\mu} ({\tilde \gamma}^a)$
being Dirac matrices in curved(flat) space-time.

Further $\psi$ are decomposed in discrete modes $k$ and spin $ (s/2)$ with $s =
\pm 1$ as
$$\psi = \sum_{s = \pm 1} \sum_{k = - \infty}^{\infty} [b_{k,s} \psi_{I k,s} +
d^{\dag}_{-k,s} { \psi}_{II k,s} ] \eqno(A12)$$
$$\psi^{\dag} = \sum_{s = \pm 1} \sum_{k = - \infty}^{\infty} [
{\bar \psi}_{I k,s} {\tilde \gamma}^0 b^{\dag}_{k,s} +
 {\bar \psi}_{II k,s}{\tilde \gamma}^0 d_{-k,s}] \eqno(A13)$$
with $b_{k,s} (b^{\dag}_{k,s})$ and $d_{-k,s} (d^{\dag}_{-k,s})$ being
annihilation (creation) operators for positive (negative) energy particles
respectively. Further, using
$$\psi_{I k,s} = f_{k,s}(\eta) e^{- i {\vec k}.{\vec x}} { u}_s,  \eqno(A14)$$
$$\psi_{II k,s} = g_{k,s}(\eta) e^{i {\vec k}.{\vec x}} {\hat u}_s,  \eqno(A15)$$
where
$$
u_1 = \begin{pmatrix}
1 \\ 0 \\ 0 \\ 0 \end{pmatrix}, u_{-1} = \begin{pmatrix}
0 \\ 1 \\ 0 \\ 0 \end{pmatrix}, {\hat u}_1 = \begin{pmatrix}
0 \\ 0 \\ 1 \\ 0 \end{pmatrix}, {\hat u}_{-1} = \begin{pmatrix}
0 \\ 0 \\ 0 \\ 1 \end{pmatrix}. 
$$

Here also, $in$- and $out$- states of $\psi$ are obtained at two extremes of
the space-time where space-time is asymptotically Minkowskian.

The Bogolubov transformations for $\psi$  are given as
\begin{eqnarray*}
b^{out}_{k,s} &=& b^{in}_{k,s} \alpha_{k,s} + d^{in {\dag}}_{-k,-s}
\beta_{k,s} \\ (b^{out}_{-k,-s})^{\dag} &=&
\alpha^*_{k,s}(b^{in}_{k,s})^{\dag} + \beta^*_{k,s} d^{in}_{-k,-s} \\
(d^{out}_{-k,-s})^{\dag} &=& b^{in}_{k,s} \alpha_{k,s} + d^{in {\dag}}_{-k,-s}
\beta_{k,s} \\ d^{out}_{-k,-s} &=& \alpha^*_{k,s}(b^{in}_{k,s})^{\dag} + \beta^*_{k,s} d^{in}_{-k,-s}.
\end{eqnarray*}

The Bogolubov coefficients $\alpha_{k,s}$ and $\beta_{k,s}$ satisfy the
condition 
$$ |\alpha_{k,s}|^2 + |\beta_{k,s}|^2 = 1 ,  \eqno(A 16)$$
where
$$ \alpha_{k,s} = \int_{t = constant} d^3x \psi^{in}_{I k,s} \psi^{out
  {\dag}}_{I k,s} \eqno(A 17)$$
and
$$ \beta_{k,s} = \int_{t = constant} d^3x \psi^{in}_{II -k,-s} \psi^{out
  {\dag}}_{II -k,-s}. \eqno(A 18)$$

\end{document}